\documentclass[onecolumn,showpacs,aps,epsfig]{revtex4}

\usepackage{graphicx}
\usepackage{epstopdf}
\usepackage{latexsym}

\def\be{\begin{equation}}
\def\ee{\end{equation}}
\def\bea{\begin{eqnarray}}
\def\eea{\end{eqnarray}}

\newcommand{\bear}{\begin{eqnarray}}

\newcommand{\eear}{\end{eqnarray}}

\newbox\pippobox

\def\be{\begin{equation}}
\def\ee{\end{equation}}
\def\bea{\begin{eqnarray}}
\def\eea{\end{eqnarray}}
\def\bx{{\bf x}}

\def\bk{{\bf k}}

\def\e{\epsilon}
\def\m{\mu}
\def\n{\nu}
\def\9{\nabla}

\def\s{\sigma}

\def\g{\gamma}
\def\z{\zeta}
\def\b{\beta}

\def\lam{\lambda}

\def\h{\eta}
\def\d{\delta}

\def\D{\Delta}
\def\p{\phi}

\def\nn{\nonumber}
\def\half{\frac12}
\def\le{\left}
\def\ri{\right}
\def\6{\partial}

\def\na{\nabla}
\def\vp{\varphi}

\def\f{\frac}
\def\ma{\mathcal}

\def\Mp{M_{pl}}
\def\tld{\tilde}
\def\0{(0)}
\def\half{\f{1}{2}}
\def\>{\rightarrow}

\begin{document}

\title{{\bf Scalar graviton in the healthy extension of Ho\v rava-Lifshitz theory}}

\author{Rong-Gen Cai$^{1,}$\footnote{Email: cairg@itp.ac.cn}}

\author{Bin Hu$^{1,}$\footnote{Email: hubin@itp.ac.cn}}

\author{Hong-Bo Zhang$^{1,}$\footnote{Email: hbzhang@itp.ac.cn}}

\affiliation{$^{1}$Key Laboratory of Frontiers in Theoretical
Physics, Institute of Theoretical Physics, Chinese Academy of
Sciences, P.O. Box 2735, Beijing 100190, China}

\date{\today}

\begin{abstract}
 In this note we study the linear dynamics of scalar graviton in a de Sitter background
in the infrared limit of the healthy extension of Ho\v rava-Lifshitz
gravity with the dynamical critical exponent $z=3$. Both our analytical and
numerical results show that the non-zero Fourier modes of scalar
graviton oscillate with an exponentially damping amplitude on the
sub-horizon scale, while on the super-horizon scale, the phases are
frozen and they approach to some asymptotic values. In addition, as
the case of the non-zero modes on super-horizon scale, the zero mode
also initially decays exponentially and then approaches to an
asymptotic constant value.
\end{abstract}

\pacs{04.60.-m; 98.80.-k; 98.80.Bp; 98.80.Cq; 98.80.Qc}

\maketitle

\section{Introduction}
Recently, a power-counting renormalizable ultraviolet (UV) complete
quantum gravity theory was proposed by Ho\v rava
\cite{Horava:2008ih,Horava:2009uw}. This theory is characterized by
the anisotropic scaling between time and space, so the complete
diffeomorphism invariance of general relativity (GR) is lost,
instead the Ho\v rava-Lifshitz (HL) gravity is invariant under the
so-called ``foliation-preserving'' diffeomorphism  ${\rm
Diff}(M,\ma{F})$. Since this theory was proposed, a great deal of
efforts have been made, including studies of cosmology
\cite{Calcagni:2009ar,Kiritsis:2009sh,Brandenberger:2009yt,Gao:2009wn,Piao:2009ax,Takahashi:2009wc,
Wang:2010an,Myung:2009sa,Myung:2009ug,Myung:2009if,Myung:2010qg,Kim:2009zn,Mukohyama:2009gg,
Mukohyama:2009zs,Mukohyama:2009mz} and black hole physics
\cite{Lu:2009em,Cai:2009pe,Cai:2009qs,Cai:2009ar,Cai:2010ud,Cai:2009ph,Lee:2009rm,Myung:2009us},
among others \cite{Chen:2009ka,Izumi:2009ry,Cai:2009hn}. Because of
the differences of diffeomorphism groups between in HL and in GR,
one expects to see some new dynamical degrees of freedom of
gravitational fields in HL gravity. In the minimal realization
\cite{Horava:2009uw}, because the lapse function $N(t)$ introduced
in the Arnowitt-Deser-Misner (ADM) formalism respects the
``projectability condition'', i.e. it is a function of time only,
the usual local Hamiltonian constraint becomes into a global one
which will not affect the local dynamics. So, the absence of local
Hamiltonian constraint leads to a new scalar degree of freedom in
addition to the usual helicity-2 polarizations of the graviton at
the linear perturbation level
\cite{Horava:2009uw,Sotiriou:2009gy,Cai:2009dx,Chen:2009jr,Chen:2009vu,Gong:2010xp,
Wang:2009azb,Huang:2010rq,Wang:2009yz,Cerioni:2010uz,Izumi:2010yn}
(see \cite{Mukohyama:2010xz} for a review on the ``projectable
model''). However, further studies on the non-linear dynamics show
that the extra scalar mode suffers several pathologies, such as
instability and strong coupling problem
\cite{Blas:2009yd,Koyama:2009hc}. A naive extension to the Ho\v
rava's original proposal, which is often refereed to the
``nonprojectable model'' in literatures
\cite{Horava:2008ih,Horava:2009uw}, is to restore the full
dependence of the lapse function on space and time $N(t,\bx)$.  In
this model the scalar graviton becomes non-dynamical, because the
equation of motion for the lapse function gives the local
Hamiltonian constraint \cite{Sotiriou:2009bx,Gao:2009ht}.
Unfortunately, the ``nonprojectable model'' also confronts several
conceptual difficulties
\cite{Charmousis:2009tc,Li:2009bg,Henneaux:2009zb,Blas:2009yd}.
Recently, a new ``projectable extension'' with gauged $U(1)$
symmetry is proposed by Ho\v rava and Melby-Thompson
\cite{Horava:2010zj}.

In this note we will investigate another extension of the HL theory,
which is called healthy extension of HL gravity in literatures
\cite{Blas:2009yd,Blas:2009qj} (see a nice review in
\cite{Blas:2010hb}). In this model, a new ingredient
$a_i\equiv\na_iN/N$, which transforms under ${\rm Diff}(M,\ma{F})$
as a spatial vector and a time scalar, is introduced into the action
and the Hamiltonian constraint becomes into the second-class one. As
a result, one extra degree of freedom should appear in the healthy
extension. A preliminary parameterized post-Newtonian study in
\cite{Blas:2009qj} shows that the extra scalar graviton is free from
ghost instabilities in some parameter regions. Furthermore, the
cosmological evolutions of metric and density scalar perturbations
in both radiation and matter dominated eras are investigated in
\cite{Kobayashi:2010eh}. They found that, although the system has
two scalar degrees of freedom, corresponding to a scalar graviton
and an adiabatic matter fluctuation, the late-time evolution of
perturbations can be sufficiently specified by the value of one
gauge-invariant variable. So, it is natural to ask how the scalar
graviton behaves in the early universe, i.e. de Sitter or
inflationary phase. In this note we study the linear dynamics of HL
scalar graviton in a flat universe with a positive cosmological
constant. Both our analytical and numerical results show that the
non-zero Fourier modes of scalar graviton oscillate with an
exponentially damping amplitude on the sub-horizon scale, while on
the super-horizon scale, the phases are frozen and amplitudes
continuously decays until they approach to their asymptotic constant
values. In addition, as the case of the non-zero modes on
super-horizon scale, the zero mode also decays exponentially with
respect to time initially, and then approaches to an asymptotic
constant value.

This note is organized as follows. In section \ref{setup}, we
briefly review the ``foliation-preserving'' gauge symmetry and the
setup of the healthy extension of Ho\v rava-Lifshitz gravity in
$3+1$ dimensions. Then we construct some gauge-invariant variables
and derive the background equations in a spatially flat universe in
section \ref{gi}. In section \ref{pert}, we present both analytical
and numerical studies on the cosmological linear perturbations.
Finally, we conclude in section \ref{con}.

\section{\label{setup}The healthy extension of ho\v rava gravity}
In this section, we firstly present the gauge transformations
compatible with the ``foliation-preserving'' diffeomorphism ${\rm
Diff}(M,\ma{F})$, then briefly review the setup of the healthy
extension of HL gravity in $3+1$ dimensions.
\subsection{Gauge symmetry}
The field contents in the healthy extension of HL gravity are
 \be\label{field}
 {\rm lapse:~} N(t,\bx);,\qquad {\rm shift:~} N^i(t,\bx)\;, \qquad {\rm 3d ~spatial ~metric:~} g_{ij}(t,\bx)\;,\ee
where we abandon the ``projectablility condition'' on lapse function
$N(t,\bx)$.  In terms of these fields we can cast the 4-dimensional
line element in the ADM formalism as
 \be\label{le_adm}
 ds^2=-N^2c^2dt^2+g_{ij}\Big(dx^i+N^idt\Big)\Big(dx^j+N^jdt\Big)\;,\ee
where $x^0=ct$ and the light speed $c$ is restored explicitly in order to obtain the nonrelativistic theory from relativistic
one by taking $c\>\infty$ limit.

In the Ho\v rava's proposal the local Lorentz invariance is violated
due to the anisotropic scaling between space and time. For instance,
in $3+1$ dimensions the coordinates $(t,\bx)$ scale as
 \be\label{scaling}  t\to
\ell^z~t\;,\qquad \bx \to \ell~\bx\;,\ee
 where $z$
is called dynamical critical exponent. In the case of general $z$, the classical scaling dimensions of the fields are
 \be\label{dim_coun}
 [N^i]=[c]=\f{[dx]}{[dt]}=z-1\;,\qquad
 [g_{ij}]=[N]=0\;.\ee

In order to investigate the spacetime diffeomorphisms in the nonrelativistic theory,
it is convenient to start with the relativistic metric $g_{\m\n}$ but with $c$ restored
 \bea\label{metric1}
 g_{\m\n}&=&\le(
 \begin{array}{cc}
  -N^2+N_iN^i/c^2\;,&N_i/c\\
  N_i/c\;,&g_{ij}\\
 \end{array}
 \ri)\;,\\
 g^{\m\n}&=&\le(
 \begin{array}{cc}
  -1/N^2\;,&N^iN^{-2}/c\\
  N^iN^{-2}/c\;,&g^{ij}-N^iN^jN^{-2}/c^2\\
 \end{array}
 \ri)\;,
 \eea
where we only keep the leading terms in $1/c$ expansion.
Correspondingly, the covariant generators $\tld x^{\m}=x^{\m}+\xi^{\m}$ of spacetime diffeomorphisms are also expanded
formally with respect to the small parameter $1/c$
 \be\label{gen1}
 \xi^0=cf(t)+\ma{O}(1/c)\;,\qquad \xi^i=\z^i(t,\bx)+\ma{O}(1/c^2)\;.\ee
As stated above, the nonrelativistic transformation rules are easily obtained by taking
$c\>\infty$ limit of the relativistic diffeomorphisms
 \bea
 \d g_{ij}&=-&g_{jk}\na_i\z^k-g_{ik}\na_j\z^k-f\dot g_{ij}\label{rule1}\;,\\
 \d N_i&=&-\na_i\z^jN_j-\z^j\na_jN_i-\dot\z^jg_{ij}-\dot fN_i-f\dot N_i\label{rule2}\;,\\
 \d N&=&-\z^j\na_jN-\dot fN-f\dot N\;,\label{rule3}\eea
where $\na_i$ is the 3-dimensional covariant derivative compatible
with spatial metric $g_{ij}$ and the over dot  denotes the
derivative with respect to the nonrelativistic time $t$.

\subsection{Model setup}
Comparing with the naive promoting the lapse function to a spacetime field in the ``nonprojectable model''
\cite{Horava:2008ih,Horava:2009uw}, a new ingredient for constructing gauge-invariant terms in the action
is introduced in the healthy extension of HL gravity \cite{Blas:2009yd,Blas:2009qj}
 \be\label{a}
 a_i=\na_i\ln N(t,\bx)\;.\ee
In fact, the appearance of this term in the action is compulsory to
make theory be free of the pathologies, which has been mentioned in
the previous section. The key point of the healthy extension is that
once terms with this new ingredient are introduced in the action,
the Hamiltonian is no longer linear in the lapse, and the equation
of motion for $N(t,\bx)$ becomes into the second-class constraint.
By using the standard Hamiltonian analysis
\cite{Henneaux:1992ig,Kluson:2010nf}, the number of degrees of
freedom in the $D+1$ dimensional healthy HL gravity is
 \be\label{dof}
 \ma{N}=\half(\rm{dim}~\ma{P}-2\ma{C}_1-\ma{C}_2)
 =\half D(D-1)\;,\ee
where $\rm{dim}~\ma{P}=(D+1)(D+2)$ is the dimension of phase space,
$\ma{C}_1=2D$ is the number of first-class constraints, and
$\ma{C}_2=2$ is the number of second-class constraints. So, in the
$3+1$ spacetime this model exhibits an extra degrees of freedom
(dof) in addition to the usual transverse traceless helicity-2
gravitons.

Because our primary purpose is to investigate the linear dynamics of scalar perturbations on a flat universe, we write the most general form of the
action relevant to our interests
 \bea\label{act1}
 S_{H}&=&\f{\Mp^2}{2}\int dtd^3x \sqrt{g}N\Big\{\ma{L}_K-\ma{V}[g_{ij},a_i]\Big\}\;,\eea
where the kinetic term takes usual form
\be\label{kine}
 \ma{L}_K=K_{ij}K^{ij}-\lam K^2\;,\ee
with the extrinsic curvature defined as
$K_{ij}=(\dot g_{ij}-\na_iN_j-\na_jN_i)/2N$.
The most general potential up to the terms with dynamical critical exponent $z=3$ is given by
$\ma{V}[g_{ij},a_i]=\sum_{z=1}^3\ma{V}_z$,
 \bea
 \ma{V}_1&=&-g_1R-\eta a_ia^i+\s\;,\label{poten1}\\
 \ma{V}_2&=&\Mp^{-2}(g_2R^2+g_3R_{ij}R^{ij}+\h_2a_i\D a^i+\h_3R\na_ia^i)\;,\label{poten2}\\
 \ma{V}_3&=&\Mp^{-4}(g_4R\D R+g_5\na_iR_{jk}\na^iR^{jk}+\h_4a_i\D^2a^i+\h_5\D R\na_ia^i)\;,\label{poten3}\eea
where $R_{ij}$ and $R$ are the Ricci tensor and Ricci scalar,
respectively. In our convention the Laplace operator reads
$\D=\na_i\na^i$, its square and cubic are $\D^2=\D\cdot\D$ and
$\D^3=\D\cdot\D\cdot\D$. A parity-violating term
$\e^{ijk}R_{il}\na_jR^l_{~k}$ is absent in our paper because it only
affects the tensor cosmological perturbations at the linear order.
However, this term will contribute to the scalar perturbations at
the high-order level due to the appearance of coupling vertices
between scalar and tensor modes.

Now we would like to write out the equation of motion for this system of gravity.
Variation with respect to the lapse $N(t,\bx)$ gives the local Hamiltonian constraint
 \bea\label{ham}
 &&\ma{L}_K+\ma{V}+2\h\na_ia^i-\f{2\h_2}{\Mp^2}\D\na_ia^i+\f{\h_3}{\Mp^2}\D R
 -\f{2\h_4}{\Mp^4}\D^2\na_ia^i+\f{\h_5}{\Mp^4}\D^2R=0\;.\eea
Equation of motion (eom) for $N_i$ gives the ordinary momentum constraint
 \be\label{mom}
 \na_j\pi^{ij}=0\;,\ee
with $\pi^{ij}= K^{ji}-\lam Kg^{ij}$. In principle, the propagating
equations of this system can be obtained by the variation of
gravitational action with respect to $g_{ij}$, i.e. $\d S_H/\d
g_{ij}=0$. However, the explicit expressions of eom for $g_{ij}$ are
very lengthy, and one can find them in \cite{Kobayashi:2010eh}. In
the section \ref{pert}, we will derive the linear approximations of
these equations through a perturbative approach.

\section{\label{gi}gauge-invariant variables and background evolutions in a flat universe}
In this section we will firstly study the ``foliation-preserving''
gauge transformations of linear scalar modes in a flat universe, and
then demonstrate that the dynamics of scalar perturbations at the
linear level can be analyzed without complete gauge fixing due to
the novel properties of ``foliation-preserving'' gauge
transformations, finally derive the set of background equations by
using the linearized action.

The background line element of a flat Friedmann-Robertson-Walker
(FRW) universe reads
 \be\label{frw}
 ds^2=-dt^2+a^2\d_{ij}dx^idx^j\;,\ee
where the scale factor $a(t)$ is a function of time only. In our
convention we linearize the metric scalar perturbations as
 \bea\label{decom}
 \d g_{00}&=&-2\p(t,\bx)\;,\nonumber\\
 \d g_{0i}&=&a^2\6_iB\;,\nonumber\\
 \d g_{ij}&=&-2a^2(\psi\d_{ij}-\6_i\6_jE)\;,
 \eea
where $\6_i$ is the ordinary spatial derivative compatible with Kronecker delta function $\d^{ij}$ and
$\6^2=\6^i\6_i$.
The transformation rules under ``foliation-preserving'' diffeomorphism for these
modes can be obtained by virtue of (\ref{rule1})-(\ref{rule3}),
 \bea\label{rule5}
 \d\p&=&-\dot f(t)\;,\qquad \d B=-\dot\e(t,\bx)\;\qquad
 \d\psi=f(t)H(t)\;,\qquad \d E=-\e(t,\bx)\;,\eea
with $\z^i=\6_i\e+\e^i$ and $\6_i\e^i=0$.

Armed with these transformation rules, we construct some useful gauge invariant variables as follows
 \bea
 \Phi&=&\phi+(\frac{\psi}{H})^{\dot{}}\;,\label{dof2}\\
 \b&=&B-\dot{E}\;,\label{dof3}\\
 \Psi&=&\psi+H\int_{t_0}^{t}\phi d\tilde{t}\;,\label{dof4}\eea
as pointed out in \cite{Cai:2009dx}, $\Psi$ in fact is the integral version of $\Phi$.

From (\ref{rule5}), we can see that the gauge variations for the
quantities $\p(t,\bx)$ and $\psi(t,\bx)$ are two functions of time
only. This novel feature of gauge transformations allows us to
analyze the dynamics of linear perturbations by using the gauge
dependent variables, which will not lead to any gauge artifacts even
though we do not fix the gauge completely.  To see this more
clearly, here we
 give a simple example to illustrate our method.  We consider a
system consisting of two fields $\chi$ and $\g$
 \be\label{demo}
 S[\chi,\g]=-\half\int\le\{\dot\chi^2-\6_i\chi\6^i\chi-m^2\chi^2-\ma{L}^{(c)}[\chi,\g]\ri\}\;,\ee
where we explicitly write down the free sector for $\chi$, but
denote both the $\g$ sector and interaction terms by an implicit
form $\ma{L}^{(c)}$. Furthermore, we assume that the action
(\ref{demo}) is invariant under the transformation
 \bea
 \chi(t,\bx)&\>&\tld\chi(t,\bx)=\chi(t,\bx)+A(t)\;,\label{demo2}\\
 \g(t,\bx)&\>&\tld\g(t,\bx)=\g(t,\bx)+B(t)\;.\label{demo3}\eea
Under the ``gauge'' ($A=0,~B=0$), the eom for $\chi$ reads
 \be\label{demo4}
 \ddot\chi-\6^2\chi+m^2\chi+\ma{C}[\chi,\g]=0\;,\ee
with the coupling term $\ma{C}=\half\d\ma{L}^{(c)}/\d\chi$. On the
other hand, we can also write the action (\ref{demo}) in terms of
tilde quantities in (\ref{demo2}) and (\ref{demo3}) as
 \bea\label{demo5}
 \tld S[\chi,\g]&\equiv& S[\tld\chi,\tld\g]=-\half\int\le\{ \dot{\tld{\chi}}^2-\6_i\tld\chi\6^i\tld\chi-m^2\tld\chi^2-\ma{L}^{(c)}[\tld\chi,\tld\g]\ri\}\;,\\
 &=&S[\chi,\g]-\half\int 2(\dot A\dot\chi-m^2A\chi)-\half\int\tld\ma{L}^{(1)}[\chi,\g,A,B]-\half\int\ma{F}[A,B]\;,\eea
with
 \bea\label{demo6}
 \ma{L}^{(c)}[\tld\chi,\tld\g]&=&\ma{L}^{(c)}[\chi,\g]+\tld\ma{L}^{(1)}[\chi,\g,A,B]+\tld\ma{L}^{(2)}[A,B]\;,\\
 \ma{F}[A,B]&\supset&\tld\ma{L}^{(2)}[A,B]\;,\eea
where $\tld\ma{L}^{(1)}[\chi,\g,A,B]$ contains the linear terms of
$\chi$ only, while $\tld\ma{L}^{(2)}[A,B]$ and $ \ma{F}[A,B]$
contains the quadratic terms of $A,B$ only. Thus, the variation of
tilde action (\ref{demo5}) with respect to $\chi$ gives
 \be\label{demo7}
  \ddot\chi-\6^2\chi+m^2\chi+\ma{C}[\chi,\g]=\ma{S}[A,B]\;,\ee
with $\ma{S}=-\half\d\tld\ma{L}^{(1)}/\d\chi-\ddot A-m^2A$.
Comparing with the result in (\ref{demo4}), equation (\ref{demo7})
has an extra source term on the right hand side, but it  depends on
time only. Given this fact, if we turn to the Fourier space, we can
easily see that such only time dependent source term vanishes for
all $k\neq0$ modes, while it contributes a Dirac delta function to
$k=0$ mode. So, we conclude that the only time dependent gauge
transformations do not affect the dynamics of $k\neq0$ mode, i.e.
the eom for $k\neq0$ modes is the same in all gauges. These features
of Ho\v rava-type theory allow us to study the dynamics of non-zero
Fourier modes by using the gauge dependent variables $\phi$ and
$\psi$ directly without leading to any gauge artifacts.

In the rest part of this section, we list the background equations which can be obtained by varying
the linearized action with respect to the scalar perturbations. In details, eom for
$\phi$ gives Hamiltonian constraint
 \be\label{ham0}
 \s+3(1-3\lam)H^2=0\;,\ee
eom for $\psi$ or $\6^2E$ gives the evolution equation
 \be\label{evol0}
 \dot H=0\;.\ee
Finally, the momentum constraint, i.e. eom for $\6_iB$, is trivially
satisfied on the background. Note that (\ref{ham0}) and
(\ref{evol0}) give a de Sitter solution.

\section{\label{pert}Dynamics of scalar modes in linear cosmological perturbations}
In this section we will investigate the linear dynamics of scalar perturbations without any gauge fixing.
First, we have to derive the second order action
 \bea\label{grav2}
 S_H^{(2)}&=&\f{\Mp^2}{2}\int dtd^3x~a^3\Big\{(1-3\lam)\Big[3(\dot\psi+H\p)^2+2(\dot\psi+H\p)\6^2(B-\dot E)\Big]\nn\\
 &&+(1-\lam)\6^2(B-\dot E)\cdot\6^2(B-\dot E)-\h a^{-2}\p\6^2\p+2g_1a^{-2}(2\p-\psi)\6^2\psi\nn\\
 &&-\Mp^{-2}a^{-4}\Big[(16g_2+6g_3)\6^2\psi\cdot\6^2\psi-\h_2\6^2\p\cdot\6^2\p+4\h_3\6^2\psi\cdot\6^2\p\Big]\nn\\
 &&-\Mp^{-4}a^{-6}\Big[(16g_4-6g_5)\6^2\psi\cdot\6^4\psi-\h_4\6^2\p\cdot\6^4\p+4\h_5\6^2\p\cdot\6^4\psi\Big]\Big\}\;,\eea
where we have used the background equations and dropped all surface
terms. Hamiltonian constraint is given by eom for $\p$, i.e.
$\d_{\p}S_{H}^{(2)}=0$
 \bea\label{ham1}
 &&(1-3\lam)\Big[6H^2\p+2H\6^2(B-\dot E)+6H\dot\psi\Big]-2\h a^{-2}\6^2\p+4g_1a^{-2}\6^2\psi\nn\\
 &&+2\Mp^{-2}\h_2a^{-4}\6^4\p-4\Mp^{-2}\h_3a^{-4}\6^4\psi+2\Mp^{-4}\h_4a^{-6}\6^6\p-4\Mp^{-4}\h_5a^{-6}\6^6\psi=0\;.\eea
The eom for $\6^2(B-\dot E)$ produces the momentum constraint
 \bea\label{mom1}
 (1-3\lam)(H\p+\dot\psi)+(1-\lam)\6^2(B-\dot E)=0\;.\eea
Both the Hamiltonian and momentum constraint equations are
consistent the linear approximations of (\ref{ham}) and (\ref{mom}).
Finally, the propagating equations is obtained by eom for $\psi$
 \bea\label{eompsi}
 &&(1-3\lam)\Big[-6H\6^2(B-\dot E)-2\6^2(\dot B-\ddot E)-18H^2\p-6\dot H\p-6H\dot\p-18H\dot\psi-6\ddot\psi\Big]
 +4g_1a^{-2}\6^2(\p-\psi)\nn\\
 &&-\Mp^{-2}a^{-4}\Big[2(16g_2+6g_3)\6^4\psi+4\h_3\6^4\p\Big]
 -\Mp^{-4}a^{-6}\Big[2(16g_4-6g_5)\6^6\psi+4\h_5\6^6\p\Big]=0\;.\eea

\subsection{\label{nonprop}Non-propagation of the scalar graviton in the original non-projectable version}
In order to check the method presented in the section \ref{gi}, we
need to recover the results in the original ``nonprojectable model''
by using this method.  So, we turn off all $a^i$ terms in action
(\ref{act1}) and (\ref{grav2}). For further simplification, we turn
off all operators with dimension higher than 2 in the gravity
potential (\ref{poten2}) and (\ref{poten3}), i.e. merely keep $R$ and $\s$
terms in the potential (\ref{poten1}), without losing any
information. Thus, the second order gravity action (\ref{grav2})
becomes
 \bea\label{grav2.1}
 S_H^{(2)}&=&\f{\Mp^2}{2}\int dtd^3x~a^3\Big\{(1-3\lam)\Big[3(\dot\psi+H\p)^2+2(\dot\psi+H\p)\6^2(B-\dot E)\Big]\nn\\
 &&+(1-\lam)\6^2(B-\dot E)\cdot\6^2(B-\dot E)+2g_1a^{-2}(2\p-\psi)\6^2\psi\Big\}\;.\eea
Solving the Hamiltonian and momentum constraints gives
 \bea\label{constr}
 \p&=&-\f{1}{H}\dot\psi-g_1\f{1-\lam}{1-3\lam}\f{a^{-2}}{H^2}\6^2\psi\;,\\
 \6^2\b&=&g_1\f{a^{-2}}{H}\6^2\psi\;,\eea
and plugging them back into (\ref{grav2.1}), we arrive at
 \bea\label{grav2.2}
 S_H^{(2)}&=&\f{\Mp^2}{2}\int dtd^3x~a^3\le(-4g_1a^{-2}\f{\dot\psi}{H}\6^2\psi
 -2g_1a^{-2}\psi\6^2\psi-2g_1^2\f{1-\lam}{1-3\lam}\f{a^{-4}}{H^2}\6^2\psi\cdot\6^2\psi\ri)\;.\eea
From the above expression we can see that the kinetic term
$\dot\psi^2$ is absent, so at the linear perturbation level there
does not exist the dynamical scalar graviton in the ``nonprojectable
model''. This result agrees with the observations made
in~\cite{Sotiriou:2009bx,Gao:2009ht}. Beyond the non-linear order,
however, the conclusion is changed. As shown in
\cite{Horava:2010zj}, this mode becomes a dynamical one if the terms
of the fluctuation beyond the quadratic order are taken into account
in the action.  In addition, the absence of the usual quadratic time
derivative term leads to an unacceptably strong-coupling problems on
the small scale \cite{referee}.

\subsection{Linear dynamics of scalar graviton in a de Sitter universe}
As be expected from Hamiltonian analysis (\ref{dof}), there should
exist one extra scalar graviton in the pure gravitational system for
the healthy extension of the HL theory. Now, we will study the
linear dynamics of scalar graviton in a flat de Sitter universe with
the healthy extension  of the HL gravity.

Solving the momentum and Hamiltonian constraints formally, we obtain
 \be\label{mom1.1}
 \6^2\b=-\f{1-3\lam}{1-\lam}(\dot\psi+H\p)\;,\ee
and
 \be\label{ham1.1}
 \p=\f{2\le(-\f{1-3\lam}{1-\lam}\6_t-g_1a^{-2}\6^2+\Mp^{-2}\h_3a^{-4}\6^4+\Mp^{-4}\h_5a^{-6}\6^6\ri)}
 {\le(2H^2\f{1-3\lam}{1-\lam}-\h a^{-2}\6^2+\Mp^{-2}\h_2a^{-4}\6^4+\Mp^{-4}\h_4a^{-6}\6^6\ri)}\psi\;.\ee
Firstly, we plug the momentum constraint into action (\ref{grav2}) and reduce it into a functional
of two scalar fields $\p$ and $\psi$
 \bea\label{grav2.4}
 S_H^{(2)}&=&\f{\Mp^2}{2}\int dtd^3x~a^3\le\{\f{2(1-3\lam)}{1-\lam}(\dot\psi+H\p)^2
 -\h a^{-2}\p\6^2\p+2g_1a^{-2}(2\p-\psi)\6^2\psi\ri.\nn\\
 &&-\Mp^{-2}a^{-4}\Big[(16g_2+6g_3)\6^2\psi\cdot\6^2\psi-\h_2\6^2\p\cdot\6^2\p+4\h_3\6^2\psi\cdot\6^2\p\Big]\nn\\
 &&-\Mp^{-4}a^{-6}\Big[(16g_4-6g_5)\6^2\psi\cdot\6^4\psi-\h_4\6^2\p\cdot\6^4\p+4\h_5\6^2\p\cdot\6^4\psi\Big]\Big\}\;,\eea
where $\p$ field is non-dynamical dof and its eom provides a constraint
 \bea\label{ham1.2}
 &&\f{1-3\lam}{1-\lam}4H(\dot\psi+H\p)-2\h a^{-2}\6^2\p+4g_1a^{-2}\6^2\psi+2\Mp^{-2}\h_2a^{-4}\6^4\p\nn\\
 &&-4\Mp^{-2}\h_3a^{-4}\6^4\psi+2\Mp^{-4}\h_4a^{-6}\6^6\p-4\Mp^{-4}\h_5a^{-6}\6^6\psi=0\;.\eea
The eom for $\psi$ is
 \bea\label{eompsi2}
 &&-\f{4(1-3\lam)}{1-\lam}\Big[\ddot\psi+3H(\dot\psi+H\p)+\dot H\p+H\dot\p\Big]+4g_1a^{-2}\6^2(\p-\psi)
 -\Mp^{-2}a^{-4}\Big[2(16g_2+6g_3)\6^4\psi+4\h_3\6^4\p\Big]\nn\\
 &&-\Mp^{-4}a^{-6}\Big[2(16g_4-6g_5)\6^6\psi+4\h_5\6^6\p\Big]=0\;.\eea

In the Fourier space, the above two partial differential equations
reduce into the ordinary one
 \bea\label{ham1.3}
 &&\f{1-3\lam}{1-\lam}4H(\dot\psi+H\p)+2\h a^{-2}k^2\p-4g_1a^{-2}k^2\psi+2\Mp^{-2}\h_2a^{-4}k^4\p\nn\\
 &&-4\Mp^{-2}\h_3a^{-4}k^4\psi-2\Mp^{-4}\h_4a^{-6}k^6\p+4\Mp^{-4}\h_5a^{-6}k^6\psi=0\;,\eea
and
 \bea\label{eompsi2.1}
 &&-\f{4(1-3\lam)}{1-\lam}\Big[\ddot\psi+3H(\dot\psi+H\p)+\dot H\p+H\dot\p\Big]-4g_1a^{-2}k^2(\p-\psi)\nn\\
 &&-\Mp^{-2}a^{-4}k^4\Big[2(16g_2+6g_3)\psi+4\h_3\p\Big]+\Mp^{-4}a^{-6}k^6\Big[2(16g_4-6g_5)\psi+4\h_5\p\Big]=0\;,\eea
where we have suppressed the momentum $k$ implicitly in the Fourier
modes $\p_k$ and $\psi_k$. Before solving the above differential
equations, we would like to discuss the ghost-free condition which
comes from the positiveness of kinetic term $\dot\psi^2$ in the
Lagrangian (\ref{grav2.4}).  Because the field $\phi$ is not a
dynamical one, in order to analyze the dynamics of the extra scalar
graviton easily we need to solve the Hamiltonian constraint equation
(\ref{ham1.3}) and plug it into the Fourier form of the action
(\ref{grav2.4}), then the action becomes a functional of $\psi$
only. In the following discussions on the ghost-free condition we
will take this approach. Firstly, let us focus on the ultraviolet
limit $(k\>\infty)$, in which $k^{4,6}$ terms are far more important
than $k^{0,2}$ terms. Because there are some coupling constants
$(\h_2,\h_3,\h_4,\h_5)$ of higher spatial derivative terms, he
ghost-free condition can be easily satisfied, i.e., the parameter
constraints to avoid ghost instability are rather loose, here we
will not write them down explicitly. On the other hand, the
ghost-free condition in the infrared limit $(k\>0)$ is controlled by
 the parameter $\h$ only,  one can obtain an explicit
constraint on $\h$. In the infrared limit, we can safely neglect
$\ma{O}(k^{4,6})$ terms, thus (\ref{ham1.3}) reduces into
 \be\label{ghost1}
 \phi=\f{4g_1\bar k^2\psi-\f{1-3\lam}{1-\lam}4H\dot\psi}{\f{1-3\lam}{1-\lam}4H^2+2\eta\bar k^2}
 \simeq -\f{\dot\psi}{H}+\f{(1-\lam)g_1\bar k^2\psi}{(1-3\lam)H^2}+\f{(1-\lam)\h\bar k^2\dot\psi}{2(1-3\lam)H^3}\;,\ee
where we have denoted the physical wavelenght by $\bar k=k/a$. The
relevant part for the kinetic term in the action reads
 \bea\label{ghostaction}
 S_H^{(2)}&\supset&\f{\Mp^2}{2(2\pi)^3}\int dt\int d^3k a^3\le\{\f{2(1-3\lam)}{1-\lam}|\dot\psi+H\phi|^2+\bar k^2\eta|\phi|^2+\cdots\ri\}\;.
 \eea
Here we emphasize that the Fourier components $\phi,\psi$ are complex fields, in order to keep the
action real we have the absolute value symbol for each field component, or we can also write $|\phi|^2$ as $\phi(t,\bk)\phi^{\ast}(t,\bk)$. Plugging
(\ref{ghost1}) into (\ref{ghostaction}), one can easily see that the leading term in the kinetic term reads
 \be\label{ghostfree}
 {\rm Kinetic~term}\simeq\f{\h\bar k^2}{H^2}\Big |\dot\psi\Big |^2\;,\ee
from which we can see that the ghost-free condition in the infrared
limit is $\h>0$. Our result agrees with the stability constraint in
the Minkowskian background \cite{Blas:2009qj}, but contradicts with
the one in \cite{Cerioni:2010zu} where by Cerioni and Brandenberger
find the ghost-free condition is $\h<0$. At the end of the paper, we
will have more to say on this point.

 \begin{figure}[h]
    \centering
    \includegraphics[width=11.5cm]{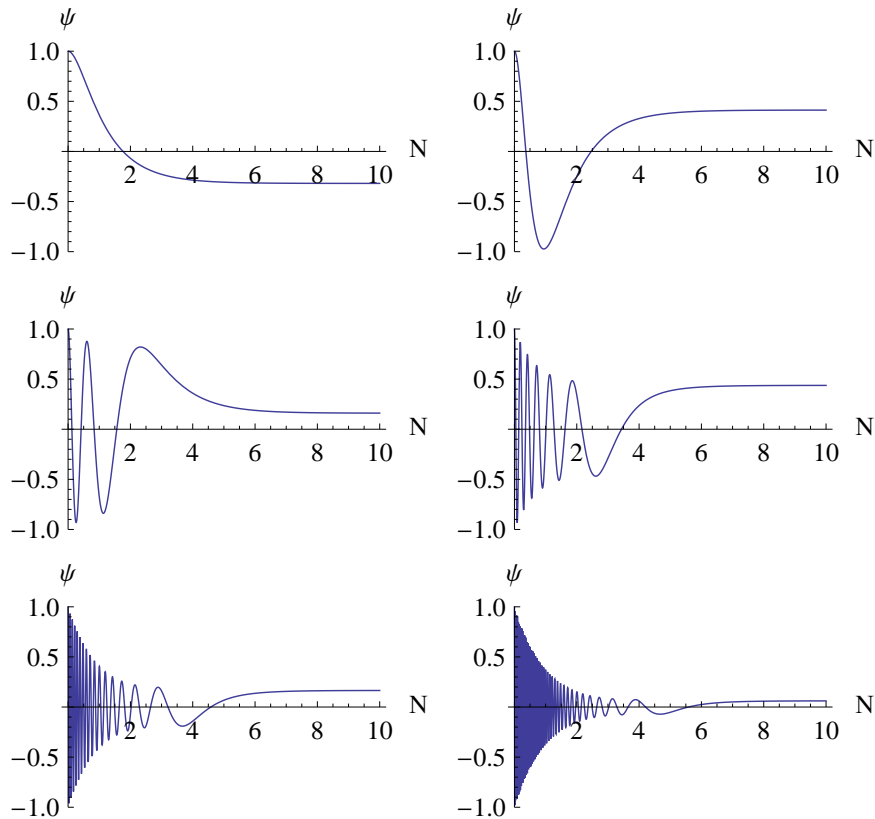}
    \caption{This figure shows the infrared behaviors of different Fourier modes of the scalar graviton $\psi$ in
    a de Sitter universe, where the horizontal axis denotes the efolding number $N=\ln (a/a_0)$. The six curves
    are the Fourier mode which exits Hubble horizon $(k_{\ast}\sim a_{\ast}H)$ at the efolding number $N_{\ast}=1$ (Top left),
    $N_{\ast}=2$ (Top right), $N_{\ast}=3$ (Middle left), $N_{\ast}=4$ (Middle right), $N_{\ast}=5$ (Bottom left),
    and $N_{\ast}=6$ (Bottom right), respectively. In
    our numerical calculations, we set the initial scale factor $a_0=1$, Planck mass $\Mp=1$, cosmological constant $\s=1$,
    gravitational coupling parameters $\lam=1.05$, $\h=0.1$ and $g_1=1$. The initial conditions are given by $\psi(t_0,\bk)=1$
    and $d\psi/dN|_{t_0,\bk}=0$. }
    \label{psieps}
 \end{figure}

Due to the cosmological interests, in the rest of this section we
will investigate both analytically and numerically the dynamics of
scalar graviton in the infrared limit, i.e. we turn off all coupling
constants $(\h_2,g_2,\h_3,g_3,\cdots)$ in the front of  the
operators with dimension larger than $2$. Firstly, we will solve the
above equations numerically with the initial condition
$\psi(t_0,\bk)=1,~d\psi/dN|_{t_0,\bk}=0$, (see Fig. \ref{psieps}).
In the numerical calculations, we plot the dynamical evolutions of
six different Fourier modes which cross horizon $(k_{\ast}\sim
a_{\ast}H)$ at the efolding number $N_{\ast}=\ln (a_{\ast}/a_0)=1$
(Top left), $N_{\ast}=2$ (Top right), $N_{\ast}=3$ (Middle left),
$N_{\ast}=4$ (Middle right), $N_{\ast}=5$ (Bottom left), and
$N_{\ast}=6$ (Bottom right), where star symbol represents for the
moment of crossing horizon and the subscript zero for the initial
time of de Sitter phase. The numerical results show that the Fourier
modes oscillate with damping amplitudes on the sub-horizon scale,
while freeze to some asymptotic values on the super-horizon scale.

Secondly, we will try to find some analytical solutions which are
able to explain the numerical behaviors. For this, it is convenient
to use the conformal time $\tau=-(aH)^{-1}$. Combining
(\ref{ham1.3}) with (\ref{eompsi2.1}), we obtain the propagating
equation for the scalar graviton in conformal time
 \be\label{eq:eompsi}
 F_{1}\left(Hk\tau\right)\psi''+F_{2}\left(Hk\tau\right)\frac{1}{\tau}\psi'+F_{3}\left(Hk\tau\right)\frac{1}{\tau^{2}}\psi=0\;,
 \ee
where prime denotes the derivative with respect to conformal time
${}'=d/d\tau$. Notice that the equation (\ref{eq:eompsi}) is
obtained non-perturbatively, and the expressions for coefficients
$F_{1}\; F_{2}$ and $F_{3}$ are given in Appendix \ref{coef}. From
the above expressions we can see that, the coefficients
$F_{1}\;,~F_{2}$ vanish if all $a_{i}$ terms are absent in the
action, i.e. $\eta_{i}=0$. So the solution of (\ref{eq:eompsi}) is
trivially $\psi=0$, which means that in this case the scalar
graviton does not exist in the case of absence of matter sources.
This result is consistent with our observations for the
``nonprojectable model'' in the subsection \ref{nonprop}.

Now, we will turn to the healthy extension model in which not all
coupling constants $\h_i$ vanish simultaneously. Because the
dimensionless quantity $k\tau$ is much less than unit on the
super-horizon region, we can use it to expand the equation
(\ref{eq:eompsi}) perturbatively, and solve them order by order. For
the super-horizon modes $(k\tau\ll1)$, the propagating equation
(\ref{eq:eompsi}) reduces to
 \begin{equation}
 \psi''+\mu k^{2}\tau\psi'+c_{s}^{2}k^{2}\psi=0\;,
 \label{eq:eompsiIR}\end{equation}
with
 \be\label{mu}
 \mu=\frac{1-\lambda}{3\lambda-1}\eta\;,\ee
and sound speed
 \be\label{cs2}
 c_{s}^{2}=\frac{\lambda-1}{3\lambda-1}\left(\frac{2g_{1}^{2}}{\eta}+g_{1}\right)\;,\ee
where we have neglected the contributions from coupling terms
$\h_2,g_2,\h_3,g_3$ etc. The dynamical stability can be satisfied
provided the sound speed is positive definitely $(c_s^2>0)$.
Considering the constraint on the coupling constant $(0<\h<2)$ from Minkowskian space \cite{Blas:2009qj},
such stability condition can be achieved in the parameter
regions $\lam\in(-\infty,1/3)\bigcup(1,+\infty)$ and $g_1\in(-\infty,-1)\bigcup(0,+\infty)$.

In order to compare with our numerical results, we rewrite the
equation (\ref{eq:eompsiIR}) in terms of the efolding number $N=\ln
(a/a_0)$
 \be\label{eompsiN}
 \f{d^2\psi}{dN^2}+\le(1-\m\f{k^2}{a^2H^2}\ri)\f{d\psi}{dN}+\f{c_s^2k^2}{a^2H^2}\psi=0\;,\ee
where the initial scale factor $a_0$ can be set to unit. For super-horizon modes, $k/aH\ll1$, the above equation reduces
into a simple form
 \be\label{eompsiN2}
 \f{d^2\psi}{dN^2}+\f{d\psi}{dN}=0\;.\ee
The general solution of the above equation reads
 \be\label{sol1}
 \psi=C_1e^{-N}+C_2\;,\ee
with two integrate constants $C_1$ and $C_2$. From the above
solution, we can see that the phases for super-horizon modes are
frozen and the amplitudes experience an exponential suppression
initially, and then approach to some asymptotic constant values.
These behaviors are in agreement with our numerical calculations.

For the sub-horizon mode $(k\tau\gg1)$, (\ref{eq:eompsi}) becomes into
 \be\label{eompsitau}
 \psi''-\f{2}{\tau}\psi'+k^2\tilde c_s^2\psi=0\;,\ee
with another sound speed
 \be\label{cs2.2}
 \tld c_s^2=\f{\lam-1}{3\lam-1}\le(\f{2g_1^2}{\h}-g_1\ri)\;.\ee
To avoid the exponential instability, the parameters must satisfy
$\lam\in(-\infty,1/3)\bigcup(1,+\infty)$ and $g_1\in(-\infty,0)\bigcup(1,+\infty)$.
The solution for (\ref{eompsitau}) reads
 \be\label{sol2}
 \psi=\f{\tau^{3/2}}{\sqrt{\tld c_sk\tau}}\le[C_1\le(\cos(\tld c_sk\tau)-\f{\sin(\tld c_sk\tau)}{\tld c_sk\tau}\ri)
 +C_2\le(\sin(\tld c_sk\tau)+\f{\cos(\tld c_sk\tau)}{\tld c_sk\tau}\ri)\ri]\;,\ee
where $C_1,~C_2$ are another two integration constants. If we
rewrite the equation (\ref{eompsitau}) with the  efolding number
 \be\label{eompsiN3}
 \f{d^2\psi}{dN^2}+3\f{d\psi}{dN}+\f{\tld c_s^2k^2}{a^2H^2}\psi=0\;.\ee
Then the general solution can be expressed as
 \be\label{sol3}
 \psi=C_1\le[\f{\tld c_sk}{aH}\cos\le(\f{\tld c_sk}{aH}\ri)-\sin\le(\f{\tld c_sk}{aH}\ri)\ri]+
 C_2\le[\cos\le(\f{\tld c_sk}{aH}\ri)+\f{\tld c_sk}{aH}\sin\le(\f{\tld c_sk}{aH}\ri)\ri]\;.\ee
From (\ref{sol3}) we can easily see that the sub-horizon Fourier
modes have two oscillating solutions. The amplitude of one solution
is large, but it is decaying with the scale factor $a^{-1}$;
the amplitude of the other is constant, but it is smaller
than the former by a factor $aH/\tilde c_sk\ll1$. The Middle right,
Bottom left and Bottom right panels in Fig. (\ref{psieps}) show that
the former exponentially suppressed modes dominate the sub-horizon
solutions in our numerical calculations.

Finally we would like to make a comment on the zero mode, for which
the above analysis becomes invalid. Fortunately, we can rewrite the
action (\ref{grav2.4}) into a gauge-invariant form, if we neglect
all spatial gradient terms by using the gauge-invariant variable
$\Psi$ (\ref{dof4})
  \bea\label{grav2.6}
  S_H^{(2)}&\simeq&\f{\Mp^2}{2}\int dtd^3x~a^3\le\{\f{2(1-3\lam)}{1-\lam}(\dot\psi+H\p)^2\ri\}\nn\\
  &=&\f{\Mp^2}{2}\int dtd^3x~a^3\le\{\f{2(1-3\lam)}{1-\lam}\dot\Psi^2\ri\}\;.\eea
Consequently, eom for $\Psi$ is given by
 \be\label{eomPsi}
 \ddot\Psi+3H\dot\Psi=0\;,\ee
with the general solution
 \be\label{gensol}
 \Psi=Ae^{-3Ht}+B\;.\ee
we see that as the case of the super-horizon non-zero modes, the
general solution of $\Psi$ also contains two parts, one of them
decays exponentially with respect to time, the other one is a
constant.

\section{\label{con}Conclusion and discussion}
In this note we investigated both analytically and numerically the
linear dynamics of scalar graviton in the healthy extension of the
HL theory in a de Sitter background.  We found that due to the
``foliation-preserving'' diffeomorphism ${\rm Diff}(M,\ma{F})$, the
gauge transformations of metric perturbations $\p$ and $\psi$ are
two functions of time only.
These novel features of gauge transformations allow us to
analyze the dynamics of linear perturbations by using the gauge
dependent variables, which will not lead to any gauge artifacts even
though we do not fix the gauge completely.
Given these
observations, we studied the linear dynamics of scalar perturbations
without any gauge fixing in section \ref{pert}. Firstly, we used our
method to study the dynamics of scalar graviton in the original
``nonprojectable model'' and found that the scalar mode does not
exist in the absence of matter sources, which is consistent with the
existing results in the literatures. Secondly, we analyzed the
dynamics of non-zero Fourier modes of scalar graviton in the infrared limit of
the healthy extension of HL gravity by our method. Both our
numerical and analytical solutions show that on the sub-horizon
scale, the non-zero Fourier modes oscillate with exponentially
damping amplitudes; but on the super-horizon scale, the phases are
frozen and amplitudes decay continuously until they approach to
their asymptotic values, i.e. the Fourier modes are conserved once
they cross the horizon. Finally, the dynamics of zero Fourier mode
of scalar graviton is also presented by using the gauge-invariant
variables. As the case of the non-zero modes on super-horizon scale,
the zero mode also decays exponentially with respect to time
initially, and then approaches to an asymptotic value.

In this note we have only investigated the dynamics of scalar
graviton in a de Sitter universe in the healthy extension of the HL
theory without matter sources.  For a more realistic case, we should
take some matter sectors into account. In particular, in the early
universe, it is natural to introduce other scalar field(s) to drive
the inflation. Thus, in such a system there will exist two scalar
perturbations at least, one is the HL scalar graviton and the others
are the fluctuations of matter fields. So, it is natural to ask
among all scalar perturbations which one is responsible for the
generation of primordial adiabatic perturbations. Some
other cosmological aspects, such as primordial non-Gaussianities and
trans-planckian problems, in the healthy extension of HL gravity are
also worthy to further investigate.

\emph{\bf Note added}: At the last stage of this work, two papers
\cite{Koh:2010wk,Cerioni:2010zu} appeared in the arXiv, which
contain some relevant discussions.  The authors of both
\cite{Koh:2010wk} and \cite{Cerioni:2010zu} consider a system within
the healthy extension of HL gravity coupled minimally with a
Lifshitz-like inflaton field $\vp$. In such a system, there exist
two dynamical scalar degrees of freedom on the linear perturbation
level around a cosmological background, one is the scalar graviton
$\psi$ which newly appears in the healthy extension, the other is
the inflaton fluctuation $\d \vp$. The authors in both
\cite{Koh:2010wk} and \cite{Cerioni:2010zu} found that the extra
scalar graviton couples with the matter field in a general FRW
universe. But in the infrared limit, these two modes decouple.
However, the reasons given in the two papers for the decouple
phenomenon is a little bit different. In \cite{Koh:2010wk} Koh and
Shin argued that the prefactor of kinetic term $\dot\psi^2$ vanishes
at the leading order $\ma{O}(k^0)$ when the momentum $(k\>0)$, i.e.
the extra scalar graviton becomes non-dynamical due to the absence
of kinetic term. On the other hand, Cerioni and Brandenberger
\cite{Cerioni:2010zu} performed the calculation to the next order
and they found that the kinetic term reads $\bar k^2\dot\psi^2/H^2$
at $\ma{O}(k^2)$ order, but the mass of the extra scalar graviton
becomes infinitely heavy when one recast the kinetic term into the
canonical form $(\dot\psi^2/2)$, because the mass term takes the
form of $(m_0\bar k^0+m_2\bar k^2+m_4\bar k^4+\cdots)H^2\psi^2/\bar
k^2$. Although the kinetic term does not vanish at the sub-leading
order, the scalar graviton still becomes non-dynamical due to the
infinitely heavy mass. Hence, at late times, it will decouple from
low energy physics and will not contribute to cosmological
perturbations on scales relevant to current observations. In our
present work, we merely considered a pure gravity theory and found
that the scalar graviton is still dynamical in the infrared limit.
This result is consistent with those in
\cite{Koh:2010wk,Cerioni:2010zu}, because when one turn off the
matter sector related with the inflaton $\vp$ in the Lagrangian,
both $m_0$ and $m_2$ vanish simultaneously, and the leading term
becomes $m_4\bar k^4\psi^2$. Furthermore, the kinetic term still
takes the form of $\bar k^2\dot\psi^2/H^2$, so after expressing the
Lagrangian in the canonical form the mass term of $\psi$ reads
$m_4\bar k^2H^2\psi^2$, i.e. {\em the extra scalar graviton keeps
massless in the infrared limit of the de Sitter background.} This
feature is the most important difference from the theory coupled
with scalar matter fields \cite{Koh:2010wk,Cerioni:2010zu}. 
Finally we would like to mention that the ghost free condition in 
a de Sitter background in our derivation ($\h>0$) 
are in contradiction from 
those obtained by Cerioni and Brandenberger ($\h<0$)
\cite{Cerioni:2010zu}. 
However, our result is consistent with the one from the
consideration in a Minkowskian background \cite{Blas:2009qj}.
So, it is necessary to fix this disagreement, but it is out of 
the scope of the present work.

\begin{acknowledgments}
 This work was partly supported by the
National Natural Science Foundation of China (No. 10821504, No.
10975168 and No.11035008), and partly by the Ministry of Science and
Technology of China under Grant No. 2010CB833004.
\end{acknowledgments}

\appendix
\section{\label{coef}}
In this appendix we list the coefficients in (\ref{eq:eompsi}),
which is obtained through a non-perturbative method. As a general
consideration, we keep all terms in the action (\ref{grav2.4})
 \begin{eqnarray*}
 F_{1}\left(x\right) & = & \frac{4\sigma}{3\left(1-\lambda\right)}G^{-1}\left(x\right)\left(G\left(x\right)-\frac{2\sigma M_{pl}^{2}}{3\left(1-\lambda\right)}\right)\;,\\
 F_{2}\left(x\right) & = & \frac{8\sigma}{9\left(1-\lambda\right)^{2}}G^{-1}\left(x\right)\left[3\left(\frac{1-\lambda}{M_{pl}^{2}}-2\sigma\right)\eta_{4}x^{6}+
 \left(4\sigma M_{pl}^{2}-3\left(1-\lambda\right)\right)\eta_{2}x^{4}+\left(2\sigma M_{pl}^{4}-3\left(1-\lambda\right)M_{pl}^{2}\right)\eta x^{2}\right]\;,\\
 F_{3}\left(x\right) & = & 8G^{-1}\left(x\right)\left[M_{pl}^{-6}\eta_{5}^{2}x^{12}-2M_{pl}^{-4}\eta_{3}\eta_{5}x^{10}+M_{pl}^{-2}\left(\eta_{3}^{2}-2g_{1}\eta_{5}\right)x^{8}+
 \left(2g_{1}\eta_{3}+M_{pl}^{-2}\frac{\sigma}{1-\lambda}\eta_{5}\right)x^{6}\right.\\
 &  & \left.+\left(g_{1}^{2}M_{pl}^{2}-\frac{1}{3}\frac{\sigma}{1-\lambda}\eta_{3}\right)x^{4}+\frac{1}{3}\frac{\sigma}{1-\lambda}g_{1}M_{pl}^{2}x^{2}\right]\\
 &  & +8\sigma\left(1-\lambda\right)^{-1}G^{-2}\left(x\right)\left[2M_{pl}^{-4}\eta_{5}\eta_{4}x^{12}-\left(\frac{4}{3}\eta_{5}\eta_{2}+2\eta_{3}\eta_{4}\right)M_{pl}^{-2}x^{10}\right.\\
 &  & \left.+\left(\frac{4}{3}\eta_{3}\eta_{2}-\frac{2}{3}\eta_{5}\eta-2g_{1}\eta_{4}\right)x^{8}+\left(\frac{2}{3}\eta_{3}\eta+
 \frac{4}{3}g_{1}\eta_{2}\right)M_{pl}^{2}x^{6}+\frac{2}{3}g_{1}M_{pl}^{4}\eta x^{4}\right]\\
 &  & -\frac{32g_{4}-12g_{5}}{M_{pl}^{4}}x^{6}+\frac{32g_{2}+12g_{3}}{M_{pl}^{2}}x^{4}-4g_{1}x^{2}\;,\\
 G\left(x\right) & = & \left(-\frac{\eta_{4}}{M_{pl}^{2}}x^{6}+\eta_{2}x^{4}+\eta M_{pl}^{2}x^{2}+\frac{2\sigma M_{pl}^{2}}{3\left(1-\lambda\right)}\right)\;.
 \end{eqnarray*}

\end{document}